\documentclass[aps,prl,twocolumn,amsmath,amssymb]{revtex4-1}
\usepackage{graphicx}
\usepackage{subfigure}
\usepackage{epsfig}
\usepackage{dcolumn}
\usepackage{bm}
\usepackage{color}
\usepackage{amsbsy}
\usepackage{hyperref}

\def\be{\begin{equation}}       \def\ee{\end{equation}}
\def\bea{\begin{eqnarray}}      \def\eea{\end{eqnarray}}
\def\ba{\begin{array}}
\def\ea{\end{array}}
\def\bnum{\begin{enumerate} }
\def\enum{\end{enumerate}}

\def\nn{\nonumber}

\def\=>{\Rightarrow}
\def\>{\rightarrow}

\def\eye2{Fathbb{I}}

\def\bk{{\bf k}}
\def\bh{{\bf h}}

\def\br{{\bf r}}

\renewcommand{\v}[1]{{\bf #1}}

\newcommand{\s}{{\sigma}}

\renewcommand{\>}{\rangle}

\begin{document}

\title{Pseudospin Hall effect of light in photonic crystals with Weyl points}
\author{Luyang Wang and Shao-Kai Jian}
\affiliation{Institute for Advanced Study, Tsinghua University, Beijing 100084, China}

\begin{abstract}
The study of spin-orbit coupling of photons has attracted much attention in recent years, and leads to many potential applications in optics. In the recently discovered metamaterial -- photonic crystals with Weyl points, the pseudospin-orbit coupling of photons bears a resemblance to the spin-orbit coupling. We study the pseudospin Hall effect of light in such metamaterials and find that during total reflection, the light beam experiences a transverse shift that is proportional to the pseudospin (or the monopole charge) of the Weyl point. As the spin Hall effect has inspired many works to explore the applications in optics, exploring the coupling between the orbital and pseudospin degrees of freedom of photons can also be fruitful in the field of optics, as well as in other fields where Weyl points can be realized.
\end{abstract}
	
\date{\today}
\maketitle

{\it Introduction}.---Light has both orbital and spin (polarization) degrees of freedom, and intrinsic spin-orbit coupling (SOC) which originates from Maxwell's equations\cite{Bliokh2015}. The SOC of light, due to its many possible applications in nano-optics, has drawn enormous interest in recent years. One fundamental effect of SOC is the spin Hall effect (SHE) of light\cite{Bliokh2004, Onoda2004,Onoda2006,Bliokh2006} that can be understood via the Berry curvature in Maxwell's equations; these Berry curvatures can be effectively viewed as radiated by two overlapped monopoles with opposite charges in the origin of momentum space, causing a spin-dependent anomalous velocity of the photons in an inhomogeneous medium. In particular, the light beam has a transverse shift when it is reflected or refracted in an interface between two media, and the direction of the transverse shift depends on the spin of photons\cite{Fedorov1955,Imbert1970,Aiello2008, Hosten2008,Bliokh2013,Yin2013}. This resembles the SHE of electrons in solid state materials\cite{Murakami2003, Sinova2015} with the role of the electric field replaced by the spatial inhomogeneity of the medium.

A related rapidly growing field of optics is topological photonics\cite{Lu2014}, of which the photonic crystal with Weyl points is a recent member\cite{Lu2013,Lu2015,Wang2016}. In such photonic crystals either time reversal or inversion symmetry is necessarily broken. As a result, the degeneracy between the right- and left-hand circularly polarized light is split and there is no degeneracy of photonic band except at the Weyl points, near which the motion of the photons is governed by the Weyl equation\cite{Weyl1929}. In Weyl equation, the pseudospin degrees of freedom are coupled to the orbital degrees of freedom. This pseudospin-orbit coupling (POC) of light bears a similarity to the intrinsic SOC of light, giving rise to the nontrivial pseudospin-dependent Berry curvatures. However, unlike Maxwell's equations, these emergent Weyl points in photonic crystals can possess a larger monopole charge\cite{Fang2012,Chen2016}, resulting in more exotic phenomena due to POC.

\begin{figure}
  \centering
  \subfigure[]{\includegraphics[width=5cm]{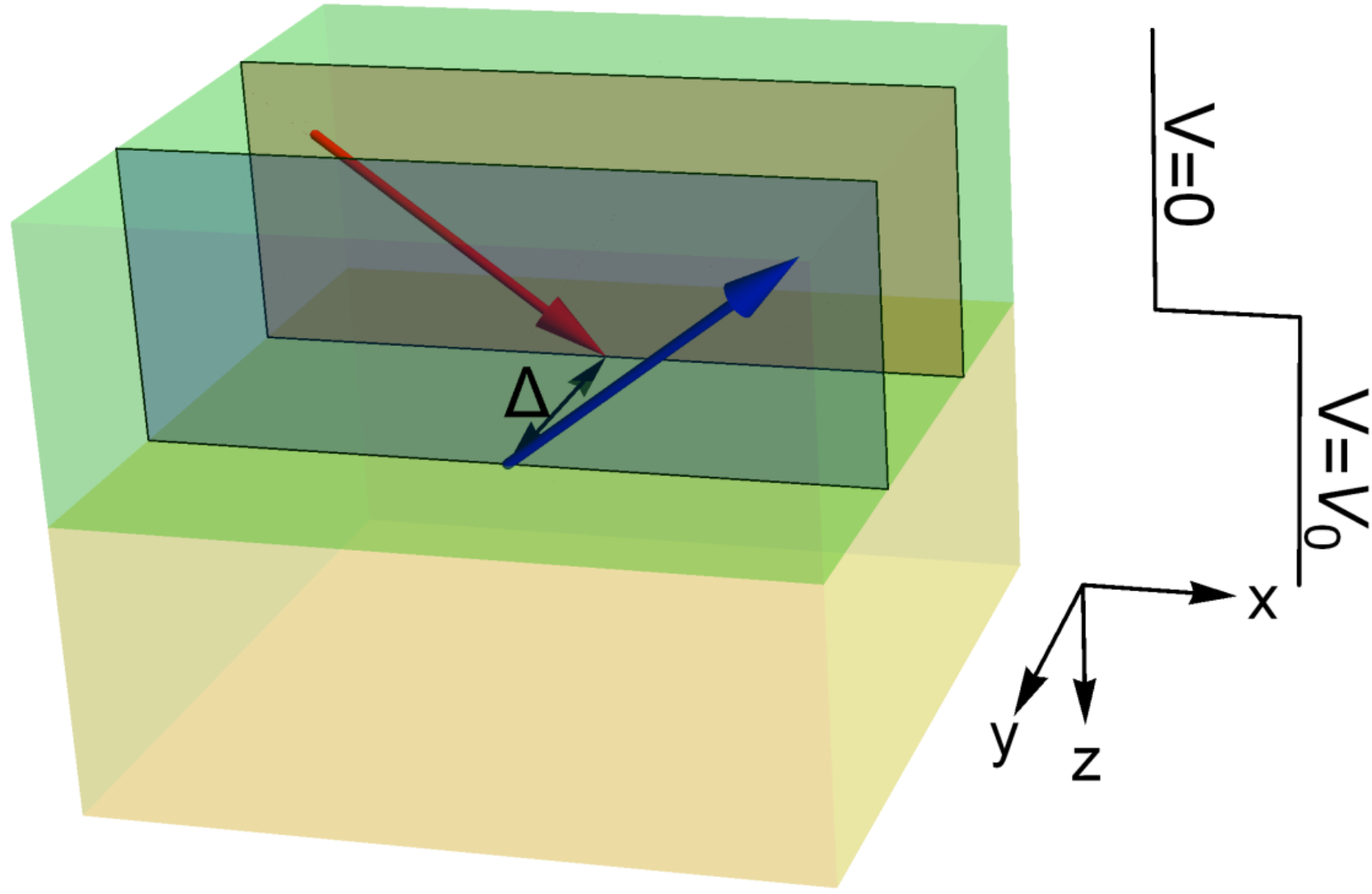}\label{schematic}}~~~~~
  \subfigure[]{\includegraphics[width=2cm]{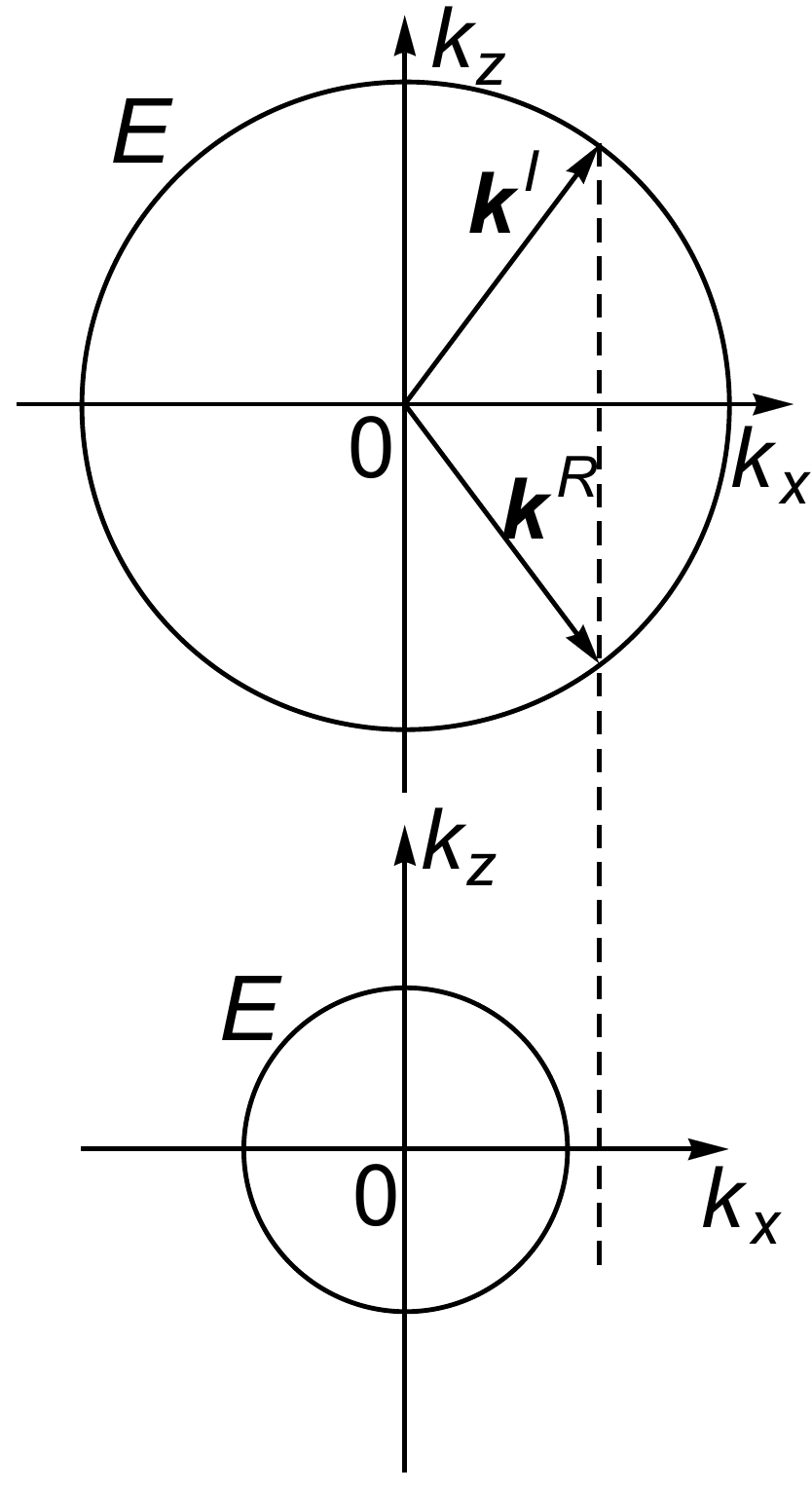}\label{tot_ref}}
  \caption{(a) Schematic figure showing a light beam incident from a photonic crystal with Weyl points to another experiences a transverse shift $\Delta$ in total reflection. (b) Schematic figure showing the condition for total reflection. The two circles represent the $k_y=0$ cross section of the equifrequency surfaces with the same frequency $E$ in the two photonic crystals, and $\bk^I$ and $\bk_R$ are the momentum of the incident and reflected photons, respectively.}
\end{figure}

In this work, we explore the pseudospin Hall effect (PHE) in photonic crystals with Weyl points: when an external force is present, either in a sharp interface as shown in Fig. \ref{schematic} or in a media with a slow varying reflection index, the photons experience a transverse shift, the direction of which depends on the pseudospin, or equivalently, the monopole charge of the Weyl point. In photonic crystals with multiple Weyl points\cite{Fang2012,Chen2016} with the monopole charge $N$ ($|N|>1$), the transverse shift is proportional to $N$. This shares some similarities with the integer quantum Hall effect, where the Hall conductance is quantized to an integer multiple of $e^2/h$. However, the transverse shift is not quantized as it depends on the frequency and the direction of the incident momentum of the photons.

In the following, we will study the PHE of a light beam occurring during a total reflection at an interface between two Weyl photonic crystals. In slightly different situations, three approaches will be used: the angular momentum conservation, the semiclassical equations of motion (EOM), and the quantum mechanical approach. Obviously, the angular momentum conservation law is only applicable in the cases with rotational symmetry. On the other hand, EOM and quantum mechanical approach are suitable for interfaces with slowly varying potential and sharp potential, respectively. The condition for total reflection is that in the photonic crystal the photons are incident on, there is no available state for the photons to transmit, as illustrated in Fig.\ref{tot_ref}.

{\it Formulation}.---Near the Weyl points of a photonic crystal, the motion of the photons is governed by the effective Hamiltonian
\begin{eqnarray}
  H_1 &=& \sum_{i=x,y,z}v_i\hat{k}_i\s_i,
\end{eqnarray}
where $v_i$ is the velocity in $i$-direction, $\hat{k}_i$ is the momentum operator and $\s_i$'s are the Pauli matrices for pseudospin degrees of freedom. The Weyl point is a monopole with topological charge $N=\pm1$ in momentum space. Since a generalization of Weyl points -- the multiple Weyl points which carry a topological charge $|N|>1$ -- have been proposed\cite{Fang2012} and realized in photonic crystals\cite{Chen2016}, we start from a more general framework by writing down the Hamiltonian of a Weyl point with an arbitrary monopole charge $N$:
\begin{eqnarray}
  H_N&=&\left\{\begin{array}{ll}
                 \left(\begin{array}{cc}\hat{k}_z&(\hat{k}_x-i\hat{k}_y)^N\\(\hat{k}_x+i\hat{k}_y)^N& -\hat{k}_z\end{array}\right)&\mbox{ for }N>0, \\
                 \left(\begin{array}{cc}\hat{k}_z&(\hat{k}_x+i\hat{k}_y)^{|N|}\\(\hat{k}_x-i\hat{k}_y)^{|N|} &-\hat{k}_z\end{array}\right)&\mbox{ for }N<0.
               \end{array}.\right.
\end{eqnarray}
When the momentum is conserved, we can replace the momentum operators by their eigenvalues.
In terms of $k_\parallel\equiv\sqrt{k_x^2+k_y^2}$ and the azimuthal angle $\phi\equiv \tan^{-1}k_y/k_x$, $H_N$ can be written in a more compact form
\begin{eqnarray}
H_N &=& \left(\begin{array}{cc}k_z&k_\parallel^{|N|}e^{-iN\phi}\\k_\parallel^{|N|}e^{iN\phi}&-k_z\end{array}\right).
\end{eqnarray}
Then the eigenfrequencies and corresponding eigenstates are
\begin{eqnarray}
  E_\pm=\pm\sqrt{k_z^2+k_\parallel^{2|N|}},~~|\psi_\pm\rangle=\frac{1}{\sqrt{1+\eta_\pm^2}} \left(\begin{array}{c}e^{-iN\phi}\\\eta_\pm\end{array}\right), \label{wavefunction}
\end{eqnarray}
where
\begin{eqnarray}
  \eta_\pm &=& \pm\sqrt{\frac{E_\pm-k_z}{E_\pm+k_z}}.
\end{eqnarray}

When an external force is exerted on the photons, the POC will give rise to the PHE. In an electronic system, the force can be exerted by applying an external electric field, and naturally exists at the interface with another electronic system. The interface between two photonic crystals can also effectively generate a force to the photons, and can be modeled by a step potential, as shown in Fig.\ref{schematic}. On the other hand, a slow variation of the envelope of the periodic refractive index $n(\br)$ also works: replacing $n(\br)$ by $n(\br)/\gamma(z)$ where $\gamma(z)$ varies slowly along $z$-direction will introduce a force in $z$-direction on the photons. Although the trajectories are distinct in the vicinity of the interface in these two situations, we will show that the transverse shifts are the same.

\begin{figure}
  \centering
  \includegraphics[width=6cm]{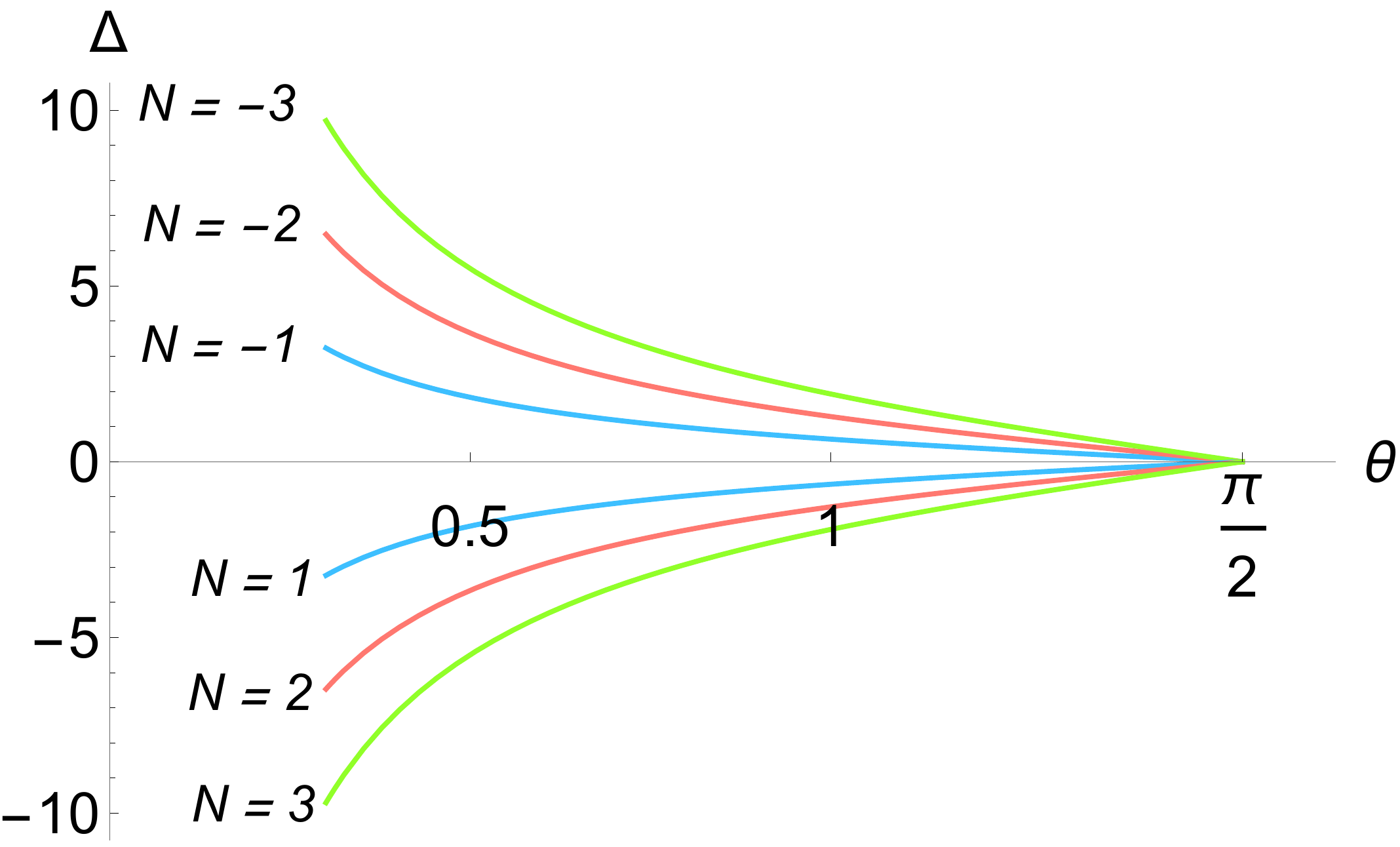}
  \caption{\label{shift}Transverse shifts of a light beam during a total reflection in the interface of two photonic crystals with the same (multiple)-Weyl points. $\theta$ is defined as $\tan \theta \equiv k_x^I/k_z^I$, where $k_i^I$ is the incident momentum. Note that it is not the incident angle in general.}
\end{figure}

{\it Conservation of TAM}.---We consider the case of total reflection of a beam of light from one photonic crystal with Weyl point of monopole charge $N$ to another one with the same monopole charge. Assume the reflection interface is at $z=0$, as shown in Fig.\ref{schematic}. Then the system has an effective continuous rotational symmetry about $z$-axis, so that the total angular momentum (TAM) $\hat{J}_z$ is conserved, where
\begin{eqnarray}
 \hat{J}_z &=& \hat{L}_z+\frac{N}{2}\s_z=-i\partial_\phi+\frac{N}{2}\s_z.
\end{eqnarray}
It is straightforward to show $[\hat{J}_z,H_N]=0$, i.e. the TAM is conserved during the reflection, $J_z^I = J_z^R $ where the index $I$ and $R$ indicate the incident and reflected photons, respectively. Semiclassically, we have
\begin{eqnarray} x^Ik_y^I-y^Ik_x^I+\frac{N}{2}\langle\s_z^I\rangle&=&x^Rk_y^R-y^Rk_x^R+\frac{N}{2}\langle\s_z^R\rangle,
\end{eqnarray}
where the average is with respect to the wave function $|\psi_\rho\rangle$ in Eq.(\ref{wavefunction}). Assume the incident beam of photons is in $xz$ plane. There is a translational symmetry in $x$ and $y$ direction, so $k_y^{I,R}=0$ and $k_x^I=k_x^R$. Then the transverse shift is
\begin{eqnarray}
\Delta=y^R-y^I=\frac{N}{2k_x^I}(\langle\s_z^R\rangle-\langle\s_z^I\rangle).
\end{eqnarray}
where
\begin{eqnarray}
\langle\s_z^{I,R}\rangle=\langle\psi_\rho|\s_z|\psi_\rho\rangle=\frac{1-\eta_\rho^2}{1+\eta_\rho^2} =\frac{k_z^{I,R}}{E},
\end{eqnarray}
and after the reflection the $z$-momentum is reversed, $k_z^R=-k_z^I$. Finally we have the transverse shift
\begin{eqnarray}
\Delta=-\frac{Nk_z^I}{E k_x^I}.\label{eq:Delta0}
\end{eqnarray}
One may incorrectly identify $k_x^I/k_z^I$ with the tangent of the incident angle. Nevertheless, unlike in a homogeneous medium, in general the velocity and the momentum of photons are not in the same direction in photonic crystals. The transverse shift is proportional to the monopole charge of the Weyl point, as shown in Fig.\ref{shift}, where $\tan \theta \equiv k_x^I/k_z^I$. In other words, the Weyl photons with different pseudospins have different transverse shifts, which is why we name the effect as pseudospin Hall effect.

For $N=\pm1$, we can easily restore the velocity parameters if the anisotropy $\delta=v_x/v_y$ is the same for both photonic crystals. Now what commutes with $H_1$ is the generalized total angular momentum
\begin{eqnarray}
J_z'=\frac{x}{\delta}k_y-\delta yk_x+\frac{N}{2}\s_z.
\end{eqnarray}
And $\langle\s_z\rangle=\frac{v_zk_z}{E}$. From the conservation of $J_z'$, we have
\begin{eqnarray}
\Delta=-\frac{v_yv_z}{v_x}\frac{Nk_z^I}{E k_x^I}.
\end{eqnarray}

The symmetry argument based on the conservation of the TAM only works if both photonic crystals have the same rotational symmetry, that is, the conserved TAM are defined in the same way. If the Weyl cones of the two photonic crystals have different pseudospins or different anisotropy, we will not be able to use this symmetry argument to derive the transverse shift.

{\it Semiclassical approach}.---The semiclassical EOM for wave packets of photons have been derived and used to study the spin Hall effect of light\cite{Onoda2004,Onoda2006}. This approach is suited to the case that the band structure varies slowly in space, and the Berry curvature has no spatial dependence. The EOM read
\begin{eqnarray}
\dot{\br}&=&\gamma(z)\nabla_\bk E_n(\bk)+\dot{\bk}\times{\bf\Omega}^{(n)}(\bk),\\
\dot{\bk}&=&(\nabla\gamma(z))E_n(\bk)\label{eq:dotk},
\end{eqnarray}
where the dots represent the time derivative, and $E_n(\bk)$ and ${\bf\Omega}^{(n)}(\bk)$ are the frequency and the Berry curvature of the $n$th band, respectively. We will ignore the index $n$ below, since in our study the wave packet stays at the same band. The dependence of the band structure on space coordinates is completely accounted for by $\gamma(z)$, so Eq.(\ref{eq:dotk}) is reduced to
\begin{eqnarray}
  \dot{k}_x=\dot{k}_y=0,\dot{k}_z &=& (\partial_z\gamma(z))E(\bk).
\end{eqnarray}
For a two band system with Hamiltonian
\begin{eqnarray}
H(\bk)=\bh(\bk)\cdot{\bm\s},
\end{eqnarray}
${\bf\Omega}(\bk)$ has a simple expression: for the upper band, the $x$-component is
\begin{eqnarray}
  \Omega_{x} &=& -\frac{\bh\cdot(\partial_{k_y}\bh\times\partial_{k_z}\bh)}{2h^3}
\end{eqnarray}
and similar for $\Omega_{y,z}$. Again, due to the rotational symmetry, we assume the incident beam is in $xz$ plane. The incident momentum is $\bk^I=(k_x^I,0,k_z^I)$, and the reflected momentum is $(k_x^I,0,-k_z^I)$. Then the band velocity in $y$-direction is $\partial_{k_y}E(\bk)=0$, and the transverse shift is from the anomalous velocity:
\begin{eqnarray}
\Delta=\int dt[\dot{\bk}\times{\bf\Omega}(\bk)]_y=\int_{k_z^I}^{-k_z^I}dk_z \Omega_x(k_x^I,0,k_z).\label{eq:Delta1}
\end{eqnarray}
We expand $(k_x+ik_y)^N$ to the first order of $k_y$,
\begin{eqnarray}
\bh(k_x,k_y,k_z)=(k_x^N, Nk_x^{N-1}k_y,k_z)+O(k_y^2),
\end{eqnarray}
from which we derive
\begin{eqnarray}
\Omega_x(k_x,0,k_z)=\frac{Nk_x^{2N-1}}{2(k_x^{2N}+k_z^2)^{\frac{3}{2}}}.
\end{eqnarray}
Plugging it into Eq.(\ref{eq:Delta1}), we have
\begin{eqnarray}
\Delta=-\frac{Nk_z^I}{k_x^I\sqrt{(k_x^I)^{2N}+k_z^{I2}}}=-\frac{Nk_z^I}{Ek_x^I},\label{eq:Delta2}
\end{eqnarray}
which agrees with the result derived by symmetry argument.

Due to Nielson-Nynomia theorem\cite{Nielsen1983}, there are always an even number of Weyl points in the bulk of periodic systems. In the simplest lattice model, the system breaks time reversal symmetry but has inversion symmetry, and there are two Weyl points in the first Brillouin zone. Assume the two Weyl points with monopole charge $\pm N$ are located at $(0,0,\pm k_z^*)$, respectively. Quantum mechanics tells us that the wave packet in one Weyl cone can either stay in the same Weyl cone or tunnel to the other Weyl cone when reflected. In the first case, in the Berry curvature $k_z$ is replaced by $k_z-k_z^*$, and the reflected momentum in $z$-direction is $2k_z^*-k_z^I$ since the frequency is conserved. Performing the integral of anomalous velocity, we find
\begin{eqnarray}
  \Delta &=& -\frac{N}{E}\frac{k_z^I-k_z^*}{k_x^I}.
\end{eqnarray}
Comparing with Eq.(\ref{eq:Delta0}) , we can just replace $k_z^I$ by $k_z^I-k_z^*$ to get this result. In the second case, the momentum of the wave packet goes from $k_z^I$ to $-k_z^I$, i.e. it passes both two Weyl points. The Berry curvatures from both Weyl points contribute to the transverse shift:
\begin{eqnarray}
\Delta&=&\int_{k_z^I}^{-k_z^I}dk_z\Omega_x\\
&=&\sum_{\rho=\pm}\int_{k_z^I}^{-k_z^I}dk_z\frac{\rho N(k_x^I)^{2N-1}}{2[(k_x^I)^{2N}+(k_z-\rho k_{z0})^2]^{\frac{3}{2}}}\\
&=&0.
\end{eqnarray}
The vanishing of transverse shift is due to the inversion symmetry and the rotational symmetry about $z$-axis: the inversion symmetry results in
\begin{eqnarray}
{\bf\Omega}(-\bk)={\bf\Omega}(\bk),
\end{eqnarray}
from which we have $\Omega_x(-k_x^I,0,-k_z)=\Omega_x(k_x^I,0,k_z)$, and the rotational symmetry gives
$\Omega_x(-k_x^I,0,k_z)=-\Omega_x(k_x^I,0,k_z)$. Combining the two symmetries, we have
$\Omega_x(k_x^I,0,-k_z)=-\Omega_x(k_x^I,0,k_z)$, which says $\Omega_x$ is an odd function of $k_z$. Thus the integral $\Delta=0$.

{\it Quantum mechanical approach}.---Quantum mechanical approach is more convenient for the case that there is a sharp interface between two photonic crystals as shown in Fig. \ref{schematic}. 
Now we consider the case that the two photonic crystals possess Weyl points with arbitrary monopole charges. The symmetry argument does not work here since the pseudospins can be different in the two crystals. In principle, the semiclassical EOM can be applied: one can interpolate the two Hamiltonians smoothly, e.g.,
\begin{eqnarray}
  H(z) &=& \frac{1}{2}(1-\tanh z)H_{N_1}+\frac{1}{2}(1+\tanh z)H_{N_2},
\end{eqnarray}
which goes to $H_{N_1}$ when $z\rightarrow-\infty$ and to $H_{N_2}$ when $z\rightarrow\infty$, and take account of phase space components of the Berry curvature, as shown in Ref.\cite{Sundaram1999}. Then one can generate the trajectories of the wave packets by numerical simulations. But the quantum mechanical approach is more convenient for this case, and can give analytically results.

We consider one (multiple) Weyl point at $\bk^*=0$ for simplicity. The Hamiltonian is
\begin{eqnarray}
  H(z) &=& \left\{\begin{array}{ll}
                 H_{N_1} &\mbox{ for }z<0, \\
                 H_{N_2}+V_0 &\mbox{ for }z>0.
               \end{array}\right.
\end{eqnarray}
The wavefunction at $z<0$ consists of the incident and reflected wavefunctions, so
\begin{eqnarray}
  \psi_1(z) &=& \frac{1}{\sqrt{1+\eta^2}}e^{ik_zz}\left(\begin{array}{c}
                                                                        e^{-iN_1\phi} \\
                                                                        \eta
                                                                      \end{array}\right)\nonumber\\
              &&+\frac{r}{\sqrt{1+\eta^2}}e^{-ik_zz}\left(\begin{array}{c}
                                                                        \eta e^{-iN_1\phi}\\
                                                                        1
                                                                      \end{array}\right) \mbox{ for }z<0,\\
  \psi_2(z) &=& \frac{t}{\sqrt{1+|\xi|^2}}e^{-\kappa z}\left(\begin{array}{c}
                                                                        \xi e^{-iN_2\phi} \\
                                                                        1
                                                                      \end{array}\right)\nn \\
               &=& \frac{t}{\sqrt{2}}e^{-\kappa z}\left(\begin{array}{c}
                                                                        e^{-i\phi_\xi-iN_2\phi}\\
                                                                        1
                                                                      \end{array}\right) ~~~~~~~\mbox{ for }z>0,
\end{eqnarray}
where we have ignored the factor $e^{i(k_xx+k_yy)}$. In the above wavefunctions, $\kappa=\sqrt{k_\parallel^{2|N_2|}-(E-V_0)^2}$, $\phi_\xi=-\tan^{-1}\frac{\kappa}{E-V_0}$ if $E>V_0$ and $\phi_\xi=-\tan^{-1}\frac{\kappa}{E-V_0}+\pi$ if $E<V_0$, and we have used the relation $\xi\equiv\frac{k_\parallel^{|N_2|}}{E-V_0-i\kappa}=\frac{k_\parallel^{|N_2|}}{\sqrt{(E-V_0)^2+\kappa^2} e^{i\phi_\xi}} =e^{-i\phi_\xi}$. Connecting the two wavefunctions at $z=0$, we have
\begin{eqnarray}
   \frac{(1+r\eta)e^{-iN_1\phi}}{\sqrt{1+\eta^2}}&=& \frac{t e^{-i\phi_\xi-iN_2\phi}}{\sqrt{2}}, \\
   \frac{\eta+r}{\sqrt{1+\eta^2}} &=& \frac{t}{\sqrt{2}}.
\end{eqnarray}
Solving for $r$, we get $r\equiv e^{i\phi_r}$ where the phase shift is
\begin{eqnarray}
  \phi_r &=& \phi_\xi-(N_1-N_2)\phi\nonumber\\
  &+&2\tan^{-1}\frac{\eta\sin(\phi_\xi-(N_1-N_2)\phi)}{1-\eta\cos(\phi_\xi-(N_1-N_2)\phi)}.
\end{eqnarray}
$\phi_\xi$ is even in $k_y$, so $\partial_{k_y}\phi_\xi|_{k_y=0}=0$. Then
\begin{eqnarray}
  \partial_{k_y}\phi_r|_{k_y=0} &=& \frac{(N_1-N_2)k_z}{(E-V_0)k_x^{N_1-N_2+1}-Ek_x}.
\end{eqnarray}
Assume the beams are Gaussian, then the center of the beam is given by the expectation value of $-i\partial_{k_y}$. Thus the center of the incident beam is $\frac{N_1\partial_{k_y}\phi|_{k_y=0}}{1+\eta^2}$, and the center of the reflected beam is $\frac{N_1\eta^2\partial_{k_y}\phi|{k_y=0}}{1+\eta^2}-\partial_{k_y}\phi_r$. The transverse shift is their difference:
\begin{eqnarray}
  \Delta =-\frac{N_1k_z}{Ek_x}+\frac{(N_1-N_2)k_z}{Ek_x-(E-V_0)k_x^{N_1-N_2+1}}.
\end{eqnarray}
Note that if $N_1=N_2=N$, we recover the results derived by symmetry argument (Eq.(\ref{eq:Delta0})) and by semiclassical EOM (Eq.(\ref{eq:Delta2})). Moreover, if the frequency of the incident photons is tuned to $E=V_0$, one can infer the monopole charge $N_2$ by measuring the transverse shift since $\Delta=-\frac{N_2k_z}{Ek_x}$ then.

{\it Discussion}.---The intense study of systems with Weyl points was initiated by the discovery of Weyl semimetals\cite{Wan2011,Lv2015a,Lv2015b,Xu2015a,Xu2015b,Yang2015NP}, in which the hypothetical relativistic paticles -- Weyl fermions -- are realized. Later it has been found that Weyl points can also exist in bosonic systems, including photonic crystals\cite{Lu2013,Lu2015,Wang2016}, magnonic systems\cite{Li2016} and acoustic systems\cite{Xiao2015}. Regardless of whether the system is fermionic or bosonic, the motion of the particles near the Weyl points are governed by the Weyl equation, which has intrinsic POC. Therefore, the PHE discussed here in the context of photonic crystals should also exist in other systems with Weyl points\cite{Jiang2015,Yang2015,Wang2017}.

The SOC of light has many potential applications in nano-optics\cite{Bliokh2015}. It can be used to generate structured optical fields, manipulate and detect small particles and control the optical wave propagation. Compared to the intrinsic SOC originated from Maxwell's equations, there are more degrees of freedom, like pseudospin and valley, in synthetic materials. And it is promising that the POC will find similar, even more applications in optics, as well as in other fields such as acoustics.

{\it Acknowledgement}. We would like to thank Hong Yao for stimulating discussions. This work was in part supported by the NSFC under Grant No. 11474175 at Tsinghua University (LW and SKJ).
\bibliography{PHEbib}

\end{document}